# Electron irradiation on multilayer $PdSe_2$ field effect transistors

A. Di Bartolomeo[1,2,a)], F. Urban[1,2,3], A. Pelella[1,2], A. Grillo[1,2], M. Passacantando[4], X. Liu[5] and F. Giubileo[2]

[1]*Department of Physics, University of Salerno, via Giovanni Paolo II, Fisciano, 84084, Italy*

[2]*CNR-SPIN, via Giovanni Paolo II, Fisciano, 84084, Italy*

[3]*INFN – Gruppo collegato di Salerno, via Giovanni Paolo II, Fisciano, 84084, Italy*

[4]*Department of Physical and Chemical Sciences, University of L'Aquila, and CNR-SPIN L'Aquila, via Vetoio, Coppito 67100, Italy*

[5]*National Laboratory of Solid State Microstructures, School of Physics, Collaborative Innovation Center of Advanced Microstructures, Nanjing University, Nanjing, 210093, China*

[a)]*Author to whom correspondence should be addressed. Electronic mail: adibartolomeo@unisa.it.*

**Abstract**

Palladium diselenide ($PdSe_2$) is a recently isolated layered material that has attracted a lot of interest for the pentagonal structure, the air stability and the electrical properties largely tunable by the number of layers. In this work, $PdSe_2$ is used in the form of multilayer as the channel of back-gate field-effect transistors, which are studied under repeated electron irradiations. Source-drain $Pd$ leads enable contacts with resistance below $350\ k\Omega \cdot \mu m$. The transistors exhibit a prevailing n-type conduction in high vacuum, which reversibly turns into ambipolar electric transport at atmospheric pressure. Irradiation by $10\ keV$ electrons suppresses the channel conductance and promptly transforms the device from n-type to p-type. An electron fluence as low as $160\ e^-/nm^2$ dramatically change the transistor behavior demonstrating a high sensitivity of $PdSe_2$ to electron irradiation. The sensitivity is lost after few exposures, that is a saturation condition is reached for fluence higher than $\sim 4000\ e^-/nm^2$. The damage induced by high electron fluence is irreversible as the device persist in the radiation-modified state for several hours, if kept in vacuum and at room temperature. With the support of numerical simulation, we explain such a behavior by electron-induced Se atom vacancy formation and charge trapping in slow trap states at the $Si/SiO_2$ interface.

---

[a)] Author to whom correspondence should be addressed. Electronic mail: adibartolomeo@unisa.it

**Introduction**

Two dimensional materials such as graphene[1,2] and transition-metal dichalcogenides[3] (TMDs) have been considered for several applications and in particular for space instrumentation[4] due to their low size and weight, the high robustness and chemical inertness, and the unique optoelectronic properties[5–8]. The use in space electronics, sensors, batteries, photovoltaics or light sources requires qualifications against vibrations and shocks, vacuum and thermal cycles, and exposure to radiation. In general, vibrations and shocks are not a threat for nanodevices and vacuum and thermal cycles are widely experimented in laboratories[9–12]. Similarly, the effect of radiation has been largely investigated for graphene[13–16] and well-known TMDs, such as $MoS_2$, $WS_2$, $MoSe_2$, and $WSe_2$[13,17–22], although most of the irradiation studies have been carried out using high energy protons, ions, electrons or $\gamma$ beams, typically in the $MeV$ range. However, the application of these materials in the context of radiation-based medical diagnostics and treatments, radioprotection, monitoring of special nuclear materials or instrumentation, and in other areas of nuclear science, also requires the understanding of their behavior when exposed to lower energy radiation sources. Low-energy (<100 $keV$) electron beams are commonly used during device fabrication by electron beam lithography (EBL) as well as for characterization and imaging through scanning and transmission electron microscopy (SEM and TEM). Exposure to low energy electrons also occurs in plasma treatments often used for device fabrication.

While high energy charged particles have reduced probability of interaction in few-layer materials and primarily damage the supporting substrate, low-energy electrons can produce significant modifications of the properties of 2D-materials based devices. Their stopping power in 2D materials becomes higher at decreasing energies below 100 $keV$[4,23]. Elastic and inelastic scattering of low energy electrons can cause ionization and atomic displacement or sputtering, creating interstitials and vacancies, which impact the electronic performance of graphene or TMDs. Indeed, using first-principles simulations, it has been calculated that the displacement threshold energies and the formation energies of chalcogen vacancies lays between 4 and 8 $eV$ in most common TMDs[21,24]. The maximum energy transferred by 80 $keV$ electrons to a $S$ (or $Mo$) atom in $MoS_2$ is 6 (or 2) $eV$[25], and the formation of vacancies due to $S$ sputtering in $MoS_2$ sheets has been demonstrated by TEM under 80 $keV$ electron beam irradiation[24]. It has been also shown that the filling of the such vacancies with impurity atoms can control the doping of the material, thereby suggesting new ways for engineering the electronic structure of TMDs[26]. Analogously, the formation of vacancies due to Se sputtering in $WSe_2$ create localized carrier-trapping deep states within the band gap and, if two adjacent Se atoms are removed in the same chalcogenide layer, the change in the local crystal structure induces a transition from direct to indirect band gap[23]. It has been also reported that $1\ keV$ to $3\ keV$ electron irradiation can transform the structure of the $MoS_2$ films from amorphous to crystalline, thus enhancing the performance of $MoS_2$- based photodetectors[27]. The capability of electron-beam irradiation to generating vacancies has enabled patterning and cutting of graphene or $MoS_2$ in nanoribbons[25,28,29]. Using SEM electron beams at energies 5, 10, 20 and $30\ keV$, it has been found that high electron fluences ($\sim 10^4\ e^-/nm^2$) result in permanent loss of photoluminesce for $WS_2$ monolayers, owing to the creation of chalcogen vacancies by knock-on damage, which cause recombination and quenching of the photoluminescence[4]. Remarkably, even at the highest fluences, the radiation induced damage was found to be mitigated if the electron energy increased to $30\ keV$ as higher-energy electrons have a smaller interaction cross-section.

In this paper, we investigate the properties of multilayer $PdSe_2$ that we use as the channel of back-gate field effect transistors. $PdSe_2$ is a noble-metal TMD with pentagonal structure and puckered layers, which is air stable[30–32]. It has been obtained in the 2D form only recently[30] and is still poorly understood, even though it has been already exploited for high-sensitivity photodetectors[33,34], field effect transistors[9,35], thermoelectric devices[36], field emitter[37] or water splitting[38]. The choice of nanosheets consisting of several layers is motivated by the recent discovery that defects in $PdSe_2$ can induce strong interlayer interactions and lead to the formation of new material phases. Indeed, it has been reported that the formation of Se vacancies by $60\ keV$ electron irradiation in $PdSe_2$ can lead to local interlayer melding and result in the formation of the new $Pd_2Se_3$ 2D phase[39]. Here, we show that the use of Pd leads over multilayer $PdSe_2$ nanosheets give rise to contacts with relatively low resistance even without any special treatments. We demonstrate that the dominant n-type conduction in high vacuum can be turned into an ambipolar or p-type one either by raising the pressure in air or by electron irradiation. We demonstrate that long exposure to $10\ keV$ electron beam in a SEM chamber changes the channel doping and transform the device from n- to p-type. Using Monte Carlo simulations, we show that the electron beam induces defects mainly in the $PdSe_2$ nanosheet and at the interface between the $Si$ back-gate and the $SiO_2$ gate dielectric. Such defects permanently change the electric conduction in the device. This study highlights a high sensitivity of $PdSe_2$ to low-energy electron irradiation, a finding that limits its use in radiation-rich environments and requires caution when an electron beam is used for device fabrication and analysis, but could enable high-sensitive radiation detectors for application in medical or nuclear instrumentation, radioprotection and radiotherapy and environmental monitoring.

**Fabrication and methods**

$PdSe_2$ nanosheets were exfoliated from bulk $PdSe_2$ single-crystal by the adhesive tape method and transferred onto a highly-doped p-type silicon substrate covered with $300\ nm$ thick $SiO_2$. The $PdSe_2$ single-crystal was synthesized through a two-step thermal process from compressed tablets of selenium (99.999 %) and palladium (99.95 %) powders, mixed in the atomic ratio of 2: 1. The tablets were heated up at 850 °C for $72\ h$ in a quartz tube at $10^{-5}\ mbar$ pressure. Finally, the cooled down poly-crystalline $PdSe_2$ tablets were mixed with $Se$ powder in a mass ratio of 1: 4 and subjected to a second similar temperature annealing cycle.

The stoichiometric $Pd$: $Se$ atomic ratio close to 1: 2 and layered crystallographic structure along the *c*-axis of the transferred nanosheets (the unit cell of $PdSe_2$ is orthorhombic with space group Pbca[30,40,41]) were confirmed by energy dispersive X-ray spectroscopy, X-ray diffractometry and Raman spectroscopy as reported in a previous work, where we used the same production batch[42].

Nanosheets with thickness of $10 - 15\ nm$ were used for device fabrication through electron-beam lithography, metal evaporation and lift-off. Several structures for transfer length measurements (TLM) were contacted using $Pd/Au$ $(5/40\ nm)$ leads. An example of a long flake, $2.5\ \mu m$ wide, with six $Pd/Au$ leads at a distance of $1.5\ \mu m$ from each other, is shown in Figure 1(a). The z-profile, obtained by atomic-force microscope (AFM), shows that the nanosheet has a thickness of 12 nm (Figure 1(b) and 1(c)) corresponding to ~30 atomic layers (the theoretical thickness of a monolayer is $0.41\ nm$[35]) .

The electrical characterization of the device was carried out by means of a Keithley 4200 semiconductor analyzer in three-terminal configuration, as shown in schematic setup of Figure 1(d), where two top metal leads are the source and drain and the silicon substrate is the back gate of a field effect transistor. The electrical measurements were performed inside a scanning electron microscope (SEM) chamber (ZEISS, LEO 1530) at room temperature and, if not otherwise specified, at pressure $\sim 10^{-6}\ torr$. The $10\ keV$ and $10\ pA$ electron beam of the SEM was used for time-controlled irradiations of the channel region.

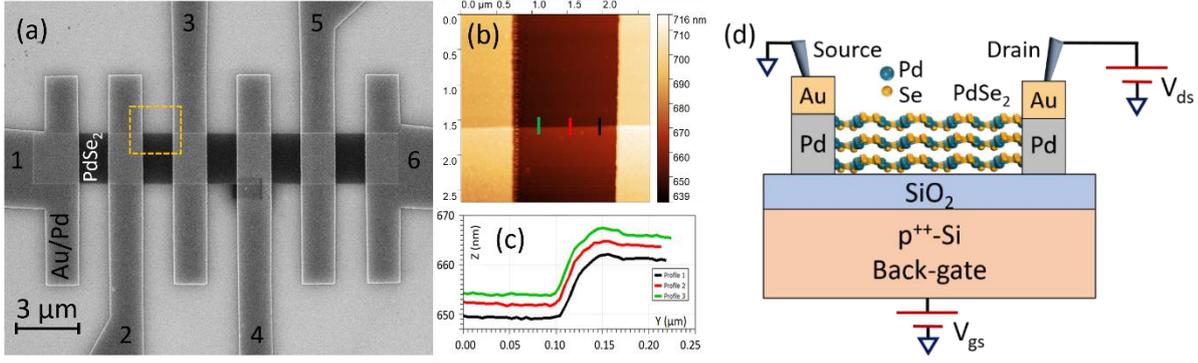

Figure 1: SEM image of a $PdSe_2$ nanosheet contacted with $Pd/Au$ leads. AFM image (b) and step height profile (c) of the $PdSe_2$ nanosheet. (d) Layout of the $PdSe_2$ field-effect transistor and biasing of the back-gate transistor in common-source configuration.

**Results and discussion**

Figure 2 (a) shows the common-source transfer characteristics (drain-current versus gate-voltage, $I_{ds} - V_{gs}$, at fixed drain-voltage, $V_{ds}$) of the transistor between metal leads labelled 5 and 6, at different pressures. The device, kept at $\sim 10^{-6}\ Torr$ for several hours, shows electron dominant conduction over the $\pm 50\ V$ range (larger biases were avoided to prevent dielectric damage). The transistor has on/off ratio around 50, a negative threshold voltage, $V_{th} \sim -10\ V$, and a maximum field-effect electron mobility $\mu = \frac{L}{W}\frac{1}{C_{SiO_2}}\frac{1}{V_{ds}}\frac{dI_{ds}}{dV_{gs}} \sim 30\ cm^2 V^{-1} s^{-1}$ ($C_{SiO_2} = 11\ nF\ cm^{-2}$ is the capacitance per unit area of the $300\ nm\ SiO_2$ gate dielectric, $L = 1.5\ \mu m$ and $W = 2.5\ \mu m$ are the channel length and width, respectively). The electron mobility is within the range reported for similar ultrathin $PdSe_2$ devices [30,33,35,43].

The low-bias output characteristics ($I_{ds} - V_{ds}$ curves at given $V_{gs}$) are straight-lines and the back-gate modulates the channel current without changing the linearity, as shown in Figure 2(b).

The n-type behavior of the transistor and the linear output curves can be understood considering a low intrinsic n-type doping of $PdSe_2$ and an ideal band alignment with the $Pd$ metal contacts[44–46]. The intrinsic n-type doping can be caused by intrinsic defects such as selenium vacancies[47–49]. The presence of defects in the $PdSe_2$ channel is strongly suggested by the clockwise hysteresis obtained when the gate voltage is swept back and forth[50–52]. Furthermore, the work function difference between the $12\ nm$ thick $PdSe_2$ (work function $\sim 5.0\ eV$[47]) and the $Pd$ contacts (work function $\geq 5.20\ eV$[53,54]) originates a band bending that favours the electron conduction, as shown in the schematic energy diagram along the channel direction in the inset of

Figure 2 (b). The low Schottky barriers at the contacts, owed to the small work function difference and the narrow bandgap of $PdSe_2$, that for $\sim 30$ layers is less than $0.3\ eV$[31,33,47], can contribute to the contact resistance but does not cause rectification at low bias. We note that, differently from here, non-linear output characteristics were reported for $PdSe_2$ with similar channel thickness and $Pd/Au$ contact[33], but this could be ascribed to a slight asymmetry of the two Schottky barriers which seems not to occur in the device under study[45].

Furthermore, Figure 2(a) shows that the behavior of the transistor is dramatically changed by pressure, as often reported for TMDs materials[42,43,55–57]. The raising pressure gradually suppresses the channel current and the transistor becomes ambipolar at atmospheric pressure. The effect of pressure has been widely investigated and mainly attributed to adsorption of molecular $O_2$ and water at defect sites which counter-dope the channel till reverting its polarity[42,48,52,58]. As shown in Figure 2(a), the effect of pressure is fully reversible as the transistor returns to its pristine status if the high vacuum is restored in a time of few hours[48].

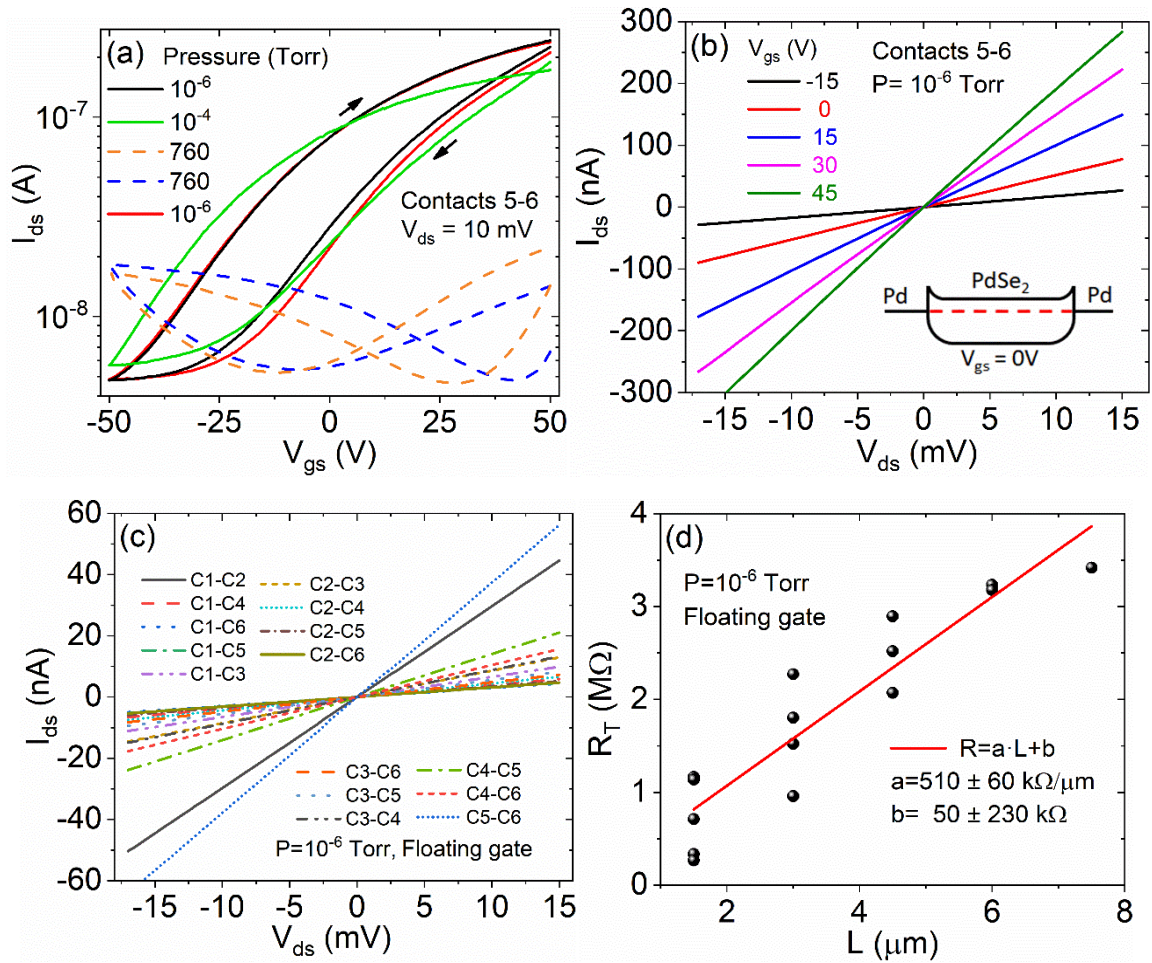

Figure 2: Transfer characteristics at various pressures in time sequence (a) and output characteristics (b) at $10^{-6}\ Torr$ of the transistor with $PdSe_2$ channel between leads 5 and 6. The inset of (b) shows the band energy diagram along the channel direction for grounded gate. Output characteristics between different combinations of the metal contacts (c) and external resistance versus channel length (L) at $\sim 10^{-6}\ Torr$ and for floating gate (the $PdSe_2$ length covered by floating metal leads between source and drain is not included in L).

Figure 2(c) displays the output characteristics at floating gate of other devices of the same TLM structure obtained considering different couples of the metal leads; remarkably, the linear behavior is maintained even if the source and drain include one or more floating contacts in between. From Figure 2(c), we extracted the external (source-drain) resistance, $R_T$, as a function of the channel length, which is shown in Figure 3(d). $R_T$ includes the two contact resistances, $R_C$, assumed equally distributed between the two contacts, and the channel resistance[59,60], $R_{Ch} = R_{Sh}\frac{W}{d}$, where $R_{Sh}$ is the channel sheet resistance in $\frac{\Omega}{square}$ and $L$ is the separation between two leads:

$$R_T = 2R_C + R_{sh}\frac{L}{W}$$

From the straight line fitting of Figure 3(d), we obtain $R_c = 25^{+120}_{-25}\ k\Omega$ and $R_{sh} = 1.27 \pm 0.15\ \frac{M\Omega}{\text{square}}$. The contact resistance can be espressed as the specific contact resistance, normalized by the width of the $PdSe_2$ channel, $\rho_c = R_c W = 60^{+300}_{-60}\ k\Omega \cdot \mu m$. Despite the high incertitude, the maximum contact resistance of 360 $k\Omega \cdot \mu m$ is significantly lower than the contact resitance $\geq 1\ M\Omega \cdot \mu m$ obtained with as-deposited Ti[35,61] or Ni[31] contacts, thus confirming the good choice of $Pd$ as for the metal leads. The sheet resistance is comparable to that measured in grain structures of monolayer $MoS_2$ with missing S[62] or undoped multilayer $MoS_2$ (with comparable mobility)[63] and even lower than in undoped $WSe_2$ monolayers[64].

To study the effect of the irradiation by an electron beam at $10\ keV$ and $10\ pA$, commonly used for SEM imaging, we selected the device between lead 1-2, which is the second most conducting transistor of the TLM structure according to Figure 2(c). The channel of such transistor had not been exposed to irradiation before, besides that from fast and low magnification imaging of the entire device area, corresponding to very low fluence on the channel region. We characterized the device shortly before and soon after each irradiation by repeatedly measuring the transfer and output characteristics. To control the electron fluence, the irradiation was performed by selecting only the channel region and imaging it for given times. Figure 3(a) shows that electron irradiation has an effect somehow similar to pressure as it suppresses the channel conductance and changes the conduction polarity. The effect of the irradiation is further clarified in Figures 3(b) and 3(c) which show the forward and the reverse branches of the transfer characteristics, respectively. Starting from the unirradiated device, three distinct groups of tranfer characteristics (taken one after the other between consecutive irradiations) can be distinguished, corresponding to successive irradiations of $10\,s$, $60\,s$, $180\,s$ (corresponding to total fluences of $165\,e^-/nm^2$, $1170\,e^-/nm^2$, $4160\,e^-/nm^2$). Surprisingly, Figure 3 (a) shows that most of the change occurs already at the low fluence of $165\,e^-/nm^2$, i.e. after the first $10\,s$ irradiation, which is enough to significantly suppress the channel conductance and cause the appearance of a p-type conduction. Such an effect is enhanced by the successive $60\,s$ irradiation. Further irradiation of $180\,s$ or more provoke only minor modifications. This observation indicates that the device becomes gradually insensitive to the electron irradiation, menaning that the effect of electron irradiation saturates after a certain fluence. However, while low fluence irradiation can make reversible changes, that might be annealed after $1\,h$ at room temperature, as we have reported before[9], here we found that a total irradiation time of $550\,s$ set the device in a new state in which it persisted for the observation time of about $5\,h$.

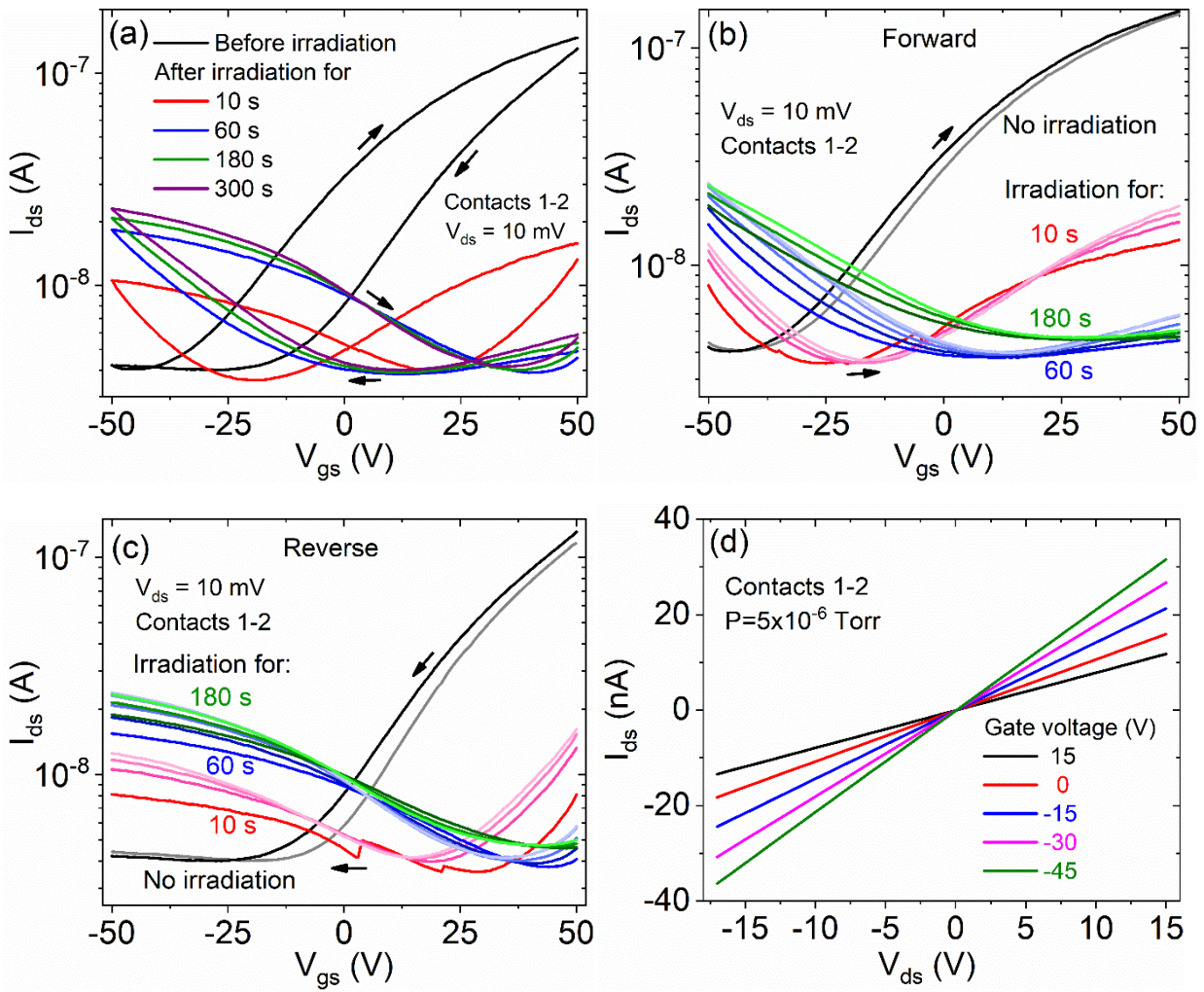

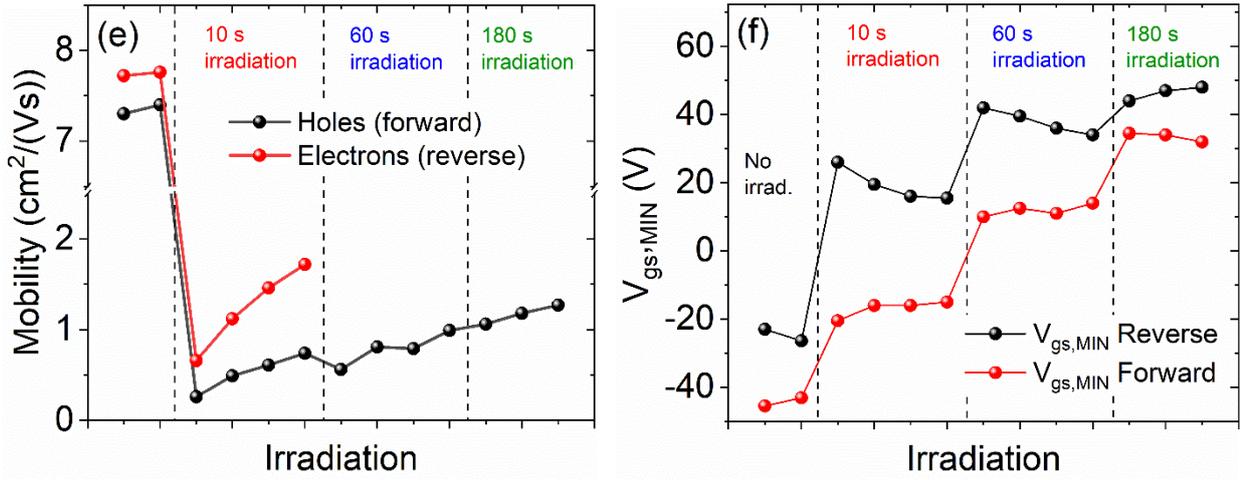

Figure 3: Full-loop (a), forward branch (b) and reverse branch (c) of the transfer characteristics of the transistor between leads 1-2 after electron beam irradiation of different duration. (d) Output characteristics after the whole cycle of irradiations. (e) Field effect mobility as a function of the electron beam irradiation. (f) Effect of the irradiation on conductance minimum, $V_{gs,MIN}$.

Figure 3(d) shows that, despite the decrease of the transistor current, the linear behaviour of the output characteristics is unaffected. The linearity is essentially related to the properties of the $Pd/PdSe_2$ interface, which is preserved as we avoided irradiation of the $Pd/PdSe_2$ contact region. Figures 3(e) and 3(f) demonstrate that the irradiation results in a significant degradation of the charge carrier mobility as well as a in a right-shift of the minimum conductance points, $V_{gs,MIN}$, obtained from both the forward and the reverse transfer curves. The mobility degradation is indicative of structural damaging and charge trapping that may generate long range coulomb scattering, while the right-shift corresponds to a doping change as the device gradually becomes p-type.

To further understand the effect of the 10 $keV$ electron beam, we performed Monte Carlo simulation to track the electrons trajectories and the energy loss into the materials. The energy released by electrons can be estimated through the cathodoluminescence, i.e. the light emission caused by electron excitation, whose intensity is shown in Figure 4 (a). We used CASINO software[65,66], a Monte Carlo simulator of electron trajectories in solids, specially designed for low-energy beam interaction in bulk and thin foils and widely used in scanning electron microscopy and microanalysis.

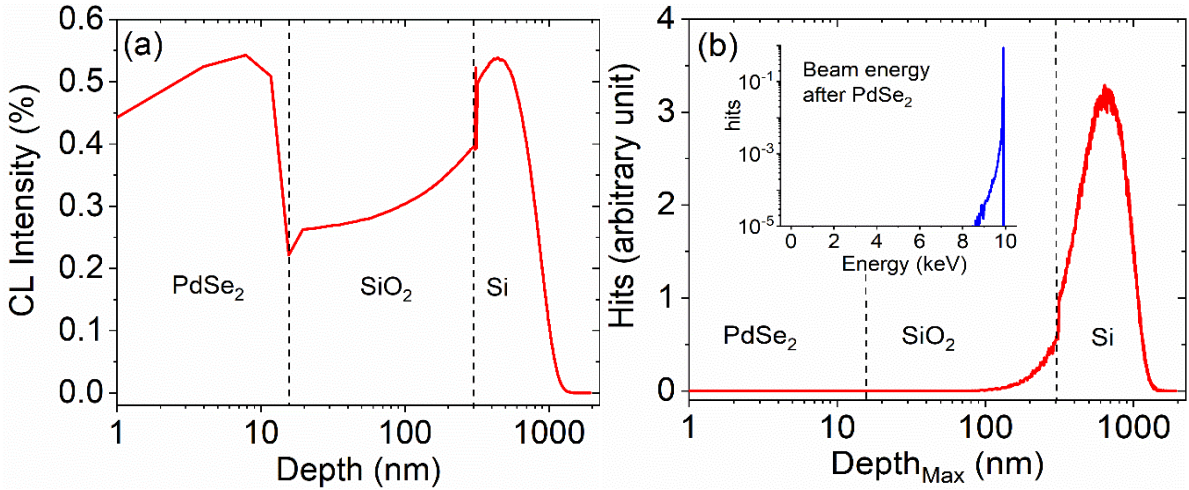

Figure 4: Monte Carlo simulation by CASINO software of the cathodoluminescence intensity (a) and the maximum penetration depth (b), normalized to the total range. The inset of figure (b) shows the transmitted energy when the 10 $keV$ beam emerges from the $PdSe_2$ nanosheet.

The simulation shows that the rate of energy loss in $PdSe_2$ and Si are comparable and higher than in $SiO_2$. As expected, most of the energy is releases in the silicon substrate where the electrons penetrate for about $\sim 1\ \mu m$ as shown by Figure 4(b) reporting the maximum penetration depth. The inset of Figure 4(b) confirms that there is energy loss in the $PdSe_2$ layer. The energy loss in $PdSe_2$ can be up to 2 $keV$, enough to cause atom displacement or sputtering[67]. The appearance of defects, preferentially Se vacancies as the energy required to displace the chalcogen ($\leq 7\ eV$) is about three times lower than that required to remove the transition metal from the same crystal[67], causes the observed reduction of the charge carrier mobility. Figure 4(a) also shows that there is an uptick in the release of energy at the $SiO_2/Si$ interface. The formation of defects at such interface introduces electron trap states[68–70], which can contribute to enhance the p-type doping of the $PdSe_2$ channel. The pile-up of negative charge at the $SiO_2/Si$ during prolonged irradiation exposures acts as an extra negatively-biased gate and right-shift $V_{gs,MIN}$ thus increasing the hole-doping of the channel. We note that the formation of defects at the $SiO_2/Si$, corresponding to trap states below the conduction band edge, created by a 10 $keV$ electron-beam exposure, has been reported in standard Si based MOS field-effect transistors[71]

**Conclusions**

We have studied $PdSe_2$ back gate transistors and shown that their electrical behavior can be dramatically affected by a SEM electron beam. We have highlighted that the beam has a twofold effect: it deteriorates the mobility of the $PdSe_2$ by creating intrinsic defects and changes the polarity of the device through pile-up of negative charges at the $SiO_2/Si$ interface. The high sensitivity of $PdSe_2$ to low energy irradiation must be taken into account and can limit the use of the material in high radiation environments as well as its treatment by electron beams or plasmas. Our study suggests the opportunity to perform electrical characterization of $PdSe_2$ devices prior to SEM imaging and demonstrates the suitability of the material for low-energy radiation detectors for medical or nuclear instrumentation as well as for environmental monitoring of caves, mines, nuclear plants or space.

**Acknowledgements**

We thank Shi Jun Liang from Nanjing University in China for providing the samples used in this study.